# Eye Gaze Controlled Robotic Arm for Persons with SSMI

Vinay Krishna Sharma a , L.R.D. Murthy a , KamalPreet Singh Saluja a , Vimal Mollyn b ,

Gourav Sharma c and Pradipta Biswas a,∗

a Indian Institute of Science, Bangalore, India

b Indian Institute of Technology, Madras, India

c Indian Institute of Information Technology, Kalyani, India

**Abstract**
**Background:** People with severe speech and motor impairment (SSMI) often uses a technique called eye pointing to communicate with outside world. One of their parents, caretakers or teachers hold a printed board in front of them and by analyzing their eye gaze manually, their intentions are interpreted. This technique is often error prone and time consuming and depends on a single caretaker.
**Objective:** We aimed to automate the eye tracking process electronically by using commercially available tablet, computer or laptop and without requiring any dedicated hardware for eye gaze tracking. The eye gaze tracker is used to develop a video see through based AR (augmented reality) display that controls a robotic device with eye gaze and deployed for a fabric printing task.
**Methodology:** We undertook a user centred design process and separately evaluated the web cam based gaze tracker and the video see through based human robot interaction involving users with SSMI. We also reported a user study on manipulating a robotic arm with webcam based eye gaze tracker.
**Results:** Using our bespoke eye gaze controlled interface, able bodied users can select one of nine regions of screen at a median of less than 2 secs and users with SSMI can do so at a median of 4 secs. Using the eye gaze controlled human-robot AR display, users with SSMI could undertake representative pick and drop task at an average duration less than 15 secs and reach a randomly designated target within 60 secs using a COTS eye tracker and at an average time of 2 mins using the webcam based eye gaze tracker.
**Conclusions:** The proposed system allows users with SSMI to manipulate physical objects without any dedicated eye gaze tracker. The novelty of the system is in terms of non-invasiveness as earlier work mostly used glass based wearable trackers or head / face tracking but no other earlier work reported use of webcam based eye tracking for controlling robotic arm by users with SSMI.

## 1. Introduction

This paper presents a non-invasive eye gaze controlled robotic manipulator for people with severe motor impairment (SSMI). We have developed a video see through eye gaze controlled interface and a webcam based eye gaze estimation software and then combined them into a single system where users with SSMI can control a robotic arm using only a webcam. We followed a user centred design process and all three modules (eye gaze controlled robot interface, webcam based gaze estimator and webcam based robot control system) were separately evaluated with end users.

Eye tracking is the process of measuring either the point of gaze where one is looking or the motion of an eye relative to the head. Eye tracking is traditionally used for analyzing visual perception, eye gaze movement [Ducholowski 2018] and making visual perception models [Biswas 2009]. In recent times, eye gaze has also been used to directly control a graphical user interface. Eye gaze controlled interfaces have been used for people with SSMI, who cannot use existing computer peripherals like mice, touchpads or keyboards. These interfaces have also recently been explored for able bodied users in situational impairments like drivers and pilots, whose hands are engaged with the primary task of driving [Biswas 2018] or flying [Shree 2018].

Persons with SSMI are not capable of using their muscles voluntarily. This causes their muscles to continuously contract leading to stiffness and tightening which interferes in normal movement and speech. The main reason for spasticity is damage to the portion of brain or spinal cord that controls voluntary movements. This damage disturbs the balance of signals from the nervous system and muscles, leading to an increased activity of the muscles [Miller 1998]. Such disabilities can be caused by birth (a defect in the neural or information processing system), or by an accidental injury. People with SSMI often uses a technique called eye pointing to communicate with outside world. One of their parents, caretakers or teachers hold a printed board in front of them and by analyzing their eye gaze manually, their intentions are interpreted (figure 1). This technique is often error prone and time consuming and depends on a single caretaker. We have tried to automate this process electronically by using commercially available tablet, computer or laptop and without requiring any dedicated hardware for eye gaze tracking.

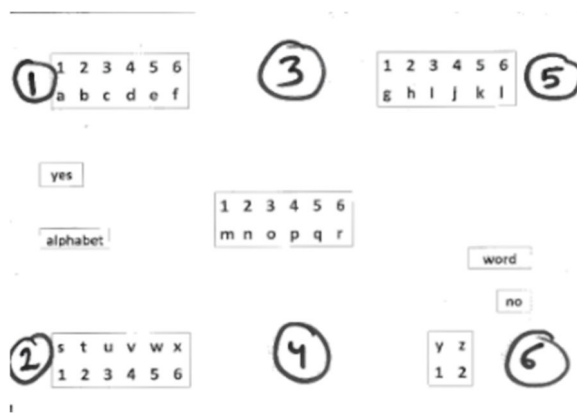

**Figure 1** Non-electronic eye pointing chart

Commercial eye gaze trackers have the advantage of higher accuracy but those need to be separately bought and configured for individual computers. Webcam based eye gaze trackers are far less accurate than infrared based commercial ones, but if it is found useful even for a limited set of applications, those can be used without the need of buying or configuring any dedicated hardware. Table 1 below compares commercial off the shelf (COTS) eye trackers with camera based ones. We have developed and compared a set of algorithms of estimating eye gaze from webcam and used it to control a human robot interface.

**Table 1** Comparing Eye Gaze Tracking Options

| | ADVANTAGES | DISADVANTAGES |
|---|---|---|
| **Cots Eye Tracker** | Accurate | Separate hardware required |
| | Low latency | Gaze controlled systems are costly |
| **Webcam Based Eye Tracker** | Low cost | Less accuracy |
| | No dedicated hardware required | Not tested for users with different range of abilities |
| | Easily configurable based on requirement or certification (automotive /aviation compliance) | |

In parallel to developing a bespoke eye gaze tracker, we took a novel approach of using a non-invasive gaze controlled system with a video see through interface to directly control a cyber-physical system like a robotic arm. In particular, we selected the use case of rehabilitating young adults with SSMI by automating a fabric printing process, which involved picking up a printing block, putting it in dry dye and subsequently placing the coloured block at a random position in a cloth. Our proposed system (figure 2) captures live video of the face of the user and process the video to estimate eye gaze direction. A transparent user interface is rendered on a video see through display and user can operate the display by dwelling eye gaze on different screen elements. As the user selects a screen element, a command will be send to an electronic interfacing circuit that will instruct the robotic arm to undertake an action fulfilling the users' intention.

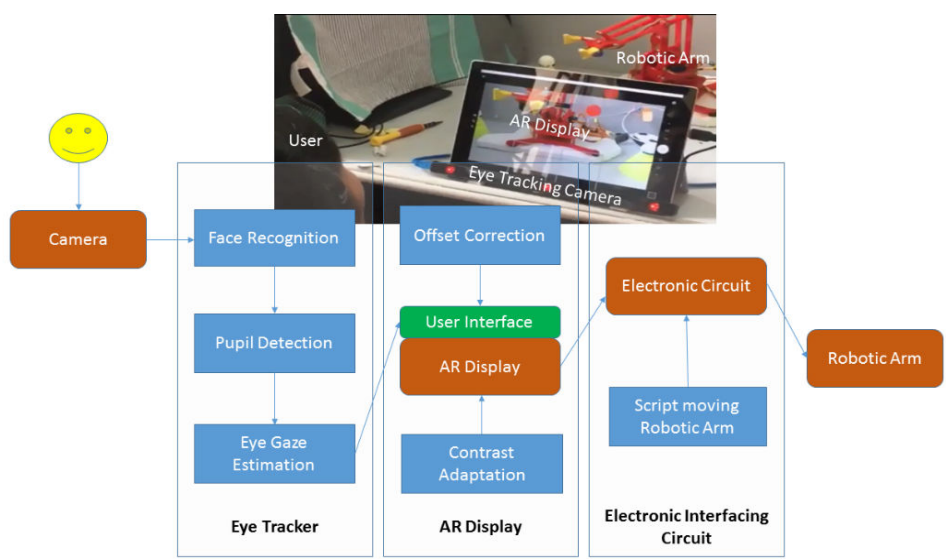

**Figure 2.** Overall Process Diagram

Main contributions of the paper are

- Developing a non-invasive eye gaze controlled video see through software interface controlling a robotic arm
- Developing webcam based eye gaze estimation algorithms
- Comparing different implementations of webcam based eye gaze tracker with users with SSMI
- Evaluating object manipulation tasks involving users with SSMI

A patent has been filed on the work at the Indian patent office with application number 201941044740

The paper is organized as follows. Section 2 presents literature survey on webcam based gaze tracking and using non-conventional modalities for human robot interaction. Our end users pose new challenge in terms of user interface design due to their different range of abilities and we described participants with SSMI in section 3. Sections 4 and 5 describe our work on developing webcam based gaze tracker and video see-through human robot interface (HRI). We also presented user studies involving both able-bodied users and people with SSMI while developing bespoke eye tracker and HRI in sections 4 and 5. Finally, section 6 presents a user study on evaluating the webcam based gaze controlled HRI followed by general discussion and conclusion at section 7.

## 2. Related Work
### 2.1. Webcam based Gaze Tracker
There is a wide variety of published research for cost effective, low resolution webcam-based eye tracking solutions. Many webcam based systems initially detect face using standard OpenCV library [Viola 2001] and then based on the relative position of pupil within the standard geometry of eyes estimate gaze position. However, none of these web cam based trackers are evaluated as extensively as commercial infra-red based gaze trackers. Khonglah [2015] reported an eye gaze tracker that used Viola-Jones [2001] detector to detect face and a blob detection algorithm to detect glint from the pupil. However, the system was tested using a heat map on interfaces having only two targets. Cristanti [2017] proposed an eye gaze controlled Android system for people with SSMI using Haar Cascade for eye detection. However, the paper did not report any evaluation on pointing and selection times for users with SSMI and accuracy was only measured in terms of eye detection. Cuong's [2014] system did not detect face, rather directly detected eyes and tested for only five positions (Right, Left, Straight, Up and Down) on screen. Sewell [2010] used a feed-forward two-layers neural network to estimate gaze vectors from the images of eyes but already reported problem in extrapolation about training the network while detecting eye gaze for one of 50 random points on screen. Kim [2015] designed a wearable system which used Emotiv Epoc EEG recording headset along with custom-built eye tracker. The pupil centre was assumed to be an ellipse and was estimated from the binarized image. RANSAC algorithm was used to remove outliers from extracted points. Furthermore, the system interpolated gaze point using second order polynomial. Eye movement was mapped to cursor movement and Brain Computer Interface (BCI) was used for selection task. The system was evaluated with four different interface protocols using Fitts' Law task with 9 able-bodied users. There are also a few commercial webcam based gaze trackers (like Web gazer, https://webgazer.cs.brown.edu/ or xLabsgaze, https://xlabsgaze.com/) but they are mainly advertised for recoding browsing behaviour of web users. Papoutsaki [2017] proposed the SearchGazer system that extends webgazer system with a regression model that maps eye features to gaze locations and search page elements during user interactions. Wakhare [2013] extracted *Between The Eyes* (BTE) features followed by eye localization using haar-like features classifier. Circular Hough transform was used to track the iris movement at real time under varying lighting conditions. Cristina [2014] suggested to detect iris from low resolution images of eye by its intensity values instead of shape followed by Kalman filters to get a smooth trajectory of cursor. Other methods like eyes region colour analysis was studied by Wojciechowski [2015] which assumed that eyelashes and iris contrast fairly with the skin colour and no external obstacles. Lin [2013] designed an approach to negate the influence of variance in lightning conditions for correctly detecting eyes. Geometrical features were used to locate the eyes correctly. The system used SVM to classify the eye images to get the gaze points at one of the 9 gaze locations. Krafka [2016] used a deep convolutional neural network trained using gazecapture data set for predicting eye gaze. Dostal's [2013] gaze controlled system is used to detect attention in one of three monitors but not used as a cursor control algorithm to operate a single display. Agrawal [2019] reported an eye gaze controlled application using viola-jones type landmark detector and a user study with able bodied users can select one of nine targets in a screen within 2.6 secs on average. There is not much reported work on using webcam based eye gaze controlled interface for users with SSMI. The ITU gaze tracker [Agustine et al, 2010] required a special hardware to hold the webcam near the eyes and it was evaluated for a typing application by able-bodied users and one motor-impaired user with 10 targets on a projected screen. Agarwal [2019] and colleagues compared a viola-jones landmark based eye gaze tracker and webgazer.js for users with SSMI but reported an average selection times more than 10 secs, which makes it unusable for practical purpose.

## 2.2. HRI with Non-conventional modalities

Use of non-traditional modalities for Human-Robot Interaction (HRI) is not a new concept although using it for people with severe disabilities remains challenging. Bannat [2009] and colleagues used direct voice input, eye gaze and soft buttons for controlling a robot in an assembly process. Alsharif [2016] and colleagues configured eye gaze movement, eye blinks and winks to the 7 degrees of freedom of a robotic arm and evaluated performance of the system with 10 participants including one person with motor impairment for a block rearrangement task. Stiefelhagen [2007] and colleagues investigated direct voice input, pointing gestures and head orientations and reported results on accuracy of each individual modality. Palinko [2016] and colleagues compared eye and head gaze based HRI for a tower building tasks and reported significant reduction in task completion time and increase in subjective preference for eye gaze tracking system compared to head tracking system. Nvidia and ITU Copenhagen [2014] published articles on gaze controlled drones achieving primitive movements by following eye gaze of an operator although the system is not yet aimed for people with different range of abilities. Bremner [2016] and colleagues investigated effect of personality cues of a robotic tele-avatar like gesture, speech and appearance on its perception by the operator. They also reported that the perception and individual behavior of operator is subjective and vary for individuals and concluded that any human robotic system should be designed considering the user background, perception and behavior. Similar findings were reported by Leite and colleagues [2012] who evaluated the empathic behavior of an autonomous social robot and reported that empathy facilitates the interaction and affects positively the perception of robot. Kohlbecher [2012] aimed to model human gaze behavior and head movement and used it for a robotic head interacting with the environment based on input from a user and reported improvement in human performance in terms of velocity and acceleration. Kim [2001] developed a real time eye tracking system and proposed its use in eye gaze controlled HRI for people with different range of abilities. The eye gaze was captured as an image by CCD cameras and processed further for head movement compensation. The setup gave accurate results within a range of 2m. However, they reported evaluation of the system for just one user with two use cases. Zaira [2014] investigated controlling robot prostheses with eye gaze control and evaluated the system with 9 able-bodied participants. Zhang [2019] used eye tracking enabled HMD to operate a remotely located robot and so far reported user studies with able–bodied participants. Chen [2014] used Electrooculography to detect eye gaze movements and used it to control robotic movement in 8-directions. The system is tested with five able-bodied participants. Fujii [2013] used Hidden Markov Models (HMMs) for gaze gesture recognition that can enable a surgeon to control a 6 DoF camera through real time gaze gestures in 2D whilst simultaneously enabling both hands to focus on the operation. Lin [2012] used an eye tracking goggles to move a wheelchair while Kuno [2003] used face direction for the same. Dziemian [2016] explored eye gaze controlled interface and a screen mounted low cost eye tracker to move a prosthetic arm but did not use video see through display and evaluated the system with only one able-bodied user. The Camera mouse system used to track nostrils for head tracking instead of eye gaze tracking and Betke [2002] reported *"There has been some success in tracking the eye, but not to the extent of determining gaze direction"*.

Use of intelligent robotics is not just limited to support people with different range of abilities but also undergone many design iterations to assist humans in workplace, factories and industries [Graser et al 2013, Peshkin et al 2001]. Such robots which physically interact with human in a co-working place sharing payload with humans are called COBOTs (COllaborative roBOTs). Based on their level of interaction, COBOTs can be classified into three categories as Co-existence – minimum interaction with no workspace sharing. Cooperation – workspace is shared with no simultaneous intervention and Collaboration- shared workspace where robot and human work simultaneously [SICK Sensor Intelligence, 2019]. Applications of cobots include complementing the skill of a human labor in manufacturing and molding by using an intuitive and adaptive graphical user interface (GUI) [Koch et al, 2017]. Cobots are also helping surgeons perform high precision surgery by optimum sensor fusion and virtual reality simulations for best surgery results [Bonneau et al, 2004]. Robots are assumed to play the roles of supervisor, operator, mechanic, peer and bystander in the era to ever-growing technological advancements [Dautenhahn, 2005]. Search and rescue robots which can be remotely operated where human and robots can work as peers. Robots can also work in proximity of humans as in Assistive Robotics where robot can be used as a tool or can act as a mentor aiding the blind, therapy to the elderly and social interaction support to autistic children [Leite et al, 2012]. Robots nowadays are being used for education and entertainment as a robotic classroom assistant, museum tour guide and a social companion. Finally, with current pace of improvements in technology and computation capabilities CoBots have found use in homes and robot maids, vacuum cleaner and robot construction [Goodrich 2008]. An inclusive CoBot can enable people with different range of abilities to undertake similar physical activities as their able bodied counterpart. In the context of state-of-the-art, we hypothesized that "A non-invasive eye gaze controlled video see through interface can enable users with SSMI to undertake representative HRI tasks as quickly as their able-bodied counterpart".

## 3. Description of Participants

Our end users with motor impairment were all school students, quadriplegic and were keen to learn operating computer. The participants were studying at The Spastic Society of India in Chennai. All trials and interactions with them were undertaken under observation by their care takers and school instructors. All necessary permissions and ethical approvals were taken before undertaking user trials. We took help from their teachers, who are rehabilitation experts, to evaluate their physical conditions. According to Gross Motor Function Classification system (GMFCS), they were all at level 5 as they could not move without wheelchair. According to Manual Ability Classification System (MACS), some of them were at level 4 and rest were at level 5. A few of them could manage to move their hand to point to a non-electronic communication chart and others only relied on eye pointing. According to Communication Function Classification System (CFCS), all of them were at level 5 as they could not speak, could make only non-speech sound and communicate only through non-electronic communication board. They did not have access to any commercially available scanning software. Their teachers and parents informed us that they were accustomed to use eye pointing with non-electronic communication chart as all of them undertakes examination using eye pointing to a non-electronic spelling chart. We have described more details on individual users in the following paragraphs.

**Participant A** | 14 years | (F) |NIOS (National Institute of Open Schooling Standard)-10$^{th}$: She understands logic easily. She has no finger isolation i.e. she cannot use hand to point something. She cannot press a switch with her hands because of tremors (athetoid movement). She has tremor in head and chin movements. She can respond yes/no by nodding head. She is comfortable with eye tracking, although sometimes it is difficult for her because of tremors.

**Participant B** | 20 years | (F) | NIOS-12$^{th}$ standard:  She takes time to understand things. She cannot hold things with her hand. She lacks finger isolation. She needs help of an assistant for keeping her head straight. She responds to yes/no by nodding her head. Her voice bottles up, but she can make some unclear sounds to draw attention. She is comfortable to use her head and chin.

**Participant C** | 15 years | (M) | NIOS-10$^{th}$ standard: He is cheerful and cooperative. He smiles and laughs when praised or when he does something well. His hands are very rigid. He has rigid body and head movement. He responds by gentle blinks to say yes.

**Participant D** | 07 years | (F) | STATE Syllabus- 3$^{rd}$ standard: She cannot speak but will respond to general talks by making voices. She is not so fluent in her hand movements but can use her hand to point something. She likes to interact with people by taking inputs from an assistant.

**Participant E** | 07 years | (M) | NIOS 2$^{nd}$ standard: He gets tensed easily. He is keen in listening to stories and loves to interact with people. He uses eyes, smile and facial expressions to respond. He cannot move his hand.

**Participant F** | 12 years | (F) | Prevocational Training: She has inhibiting nature. She is shy and tries to speak with a low voice. She gets distracted easily. She has control of her head and hand movement. She always keeps her head down.

**Participant G** | 7 years | (M) | NIOS 2$^{nd}$ standard: He is a quick learner. He can communicate well with his eyes. He can point using his hands and has good finger isolation. He can sight words on a spell chart comfortably.

**Participant H** | 20 years | (F) | NIOS 12$^{th}$ standard: She finds her interest in multimedia and can access computers well. She can use her hand to hold things like spoon. Her hands possess athetoid movement.

She can speak but with poor clarity. She can use spell chart and qwerty chart. She likes to interact with people. She has passion for making accessories. She can control head movement but has tremors.

**Participant I** | 23 years | (M) | NIOS 12th standard: He is a joyful and obedient boy. He uses spectacles but is comfortable with eye tracking. He cannot speak properly. He responds to people by making sound and moving his head. He is scared of sudden sounds. He can ride his wheel chair himself.

**Participant J** | 20 years | (M) | NIOS 12th standard: He is undergoing vocational training. He can move his chin but cannot speak. He is not able to grip or hold objects in hand because of restricted finger control and isolation. He communicates through picture chart and some gestures. He likes to interact with people.

**Participant K** | 30 years | (M) | NIOS 12th standard: He is undergoing vocation training in the institute. He has involuntary head movement and minimal chin control. He uses a spell chart to communicate and take exams. He cannot hold objects in hand due to lacking finger control and isolation. He likes to interact with people and watch TV and is able to type using keyboard with his hand.

## 4. Webcam based Gaze Tracker

In this section, we have described three different eye tracking systems and compared them through user studies

### 4.1. HoG based Gaze Tracking System

We used a pre-trained facial landmark detector with iBUG 300-W dataset [iBUG 2019], which works on classic Histogram of Oriented Gradients (HoG) feature combined with a linear classifier to detect facial landmarks [Rosebrock 2019]. In comparison, Haar cascades are a fast way to detect an object but often detect more false positives compared to HoG and linear classifier [Dalal 2005]. HoG features are capable of capturing the face or object outline/shape better than Haar features. On the other hand, simple Haar-like features can detect regions brighter or darker than their immediate surrounding region better than HoG features. In short, HoG features can describe shape better than Haar features and Haar features can describe shading better than HoG features. In this case the shape is more important as we need the landmarks of the face hence hog features produced a better result. After detecting eye region, we detected pupil location by selecting the smallest rectangle possible in the eye region where the pupil can exist. We have modified the EAR calculation formula by using the distance between the two eyes as denominator.

### 4.2. Webgazer.js

We implemented a second system using webgazer.js [Webgazer 2020; Agrawal 2019] to compare performance of the proposed system. Webgazer.js runs entirely in the client browser. WebGazer.js requires a bounding box that includes the pixels from the webcam video feed that corresponds to the detected eyes of the user. Webgazer included three external libraries (clmtracker, js_objectdetect and tracking.js) to detect face and eyes. Webgazer has methods for controlling the operation which allows us to start and stop it. We have taken the mean of last thirty points from webgazer.js for better target prediction and accuracy of system. We also calculated the mean value during this time to predict the gaze location on a webpage.

### 4.3. Intelligent System

We have developed a gaze block estimator which maps user's eye movements to 9 screen blocks using OpenFace [Baltrusaitis 2018] toolkit. Since the OpenFace (figure 3) was reported to have an error of 6° for gaze point estimation, we designed a calibration routine which uses the gaze vector data from OpenFace and maps user's eye movements to screen blocks, instead of screen points. We have divided the screen into 9 blocks of equal area. We designed a smooth-pursuit based calibration routine where a marker traverse across all these 9 blocks and user was asked to follow the marker's movement. The corresponding gaze vectors from OpenFace were recorded and stored with the respective block number as the label. Once the marker completes its path, a neural network is trained to map these gaze vectors to 9 blocks of the screen. For this classification task, we used a 2 hidden layer network with 256 and 128 neurons respectively with cross-entropy loss function with Adam optimizer. We used the 70% of the data we recorded during calibration for training, 15% for validation and the rest for testing. On a i7 processor computer, we observed that each epoch takes around 0.8 seconds and we trained the network till the test accuracy reaches 90%.

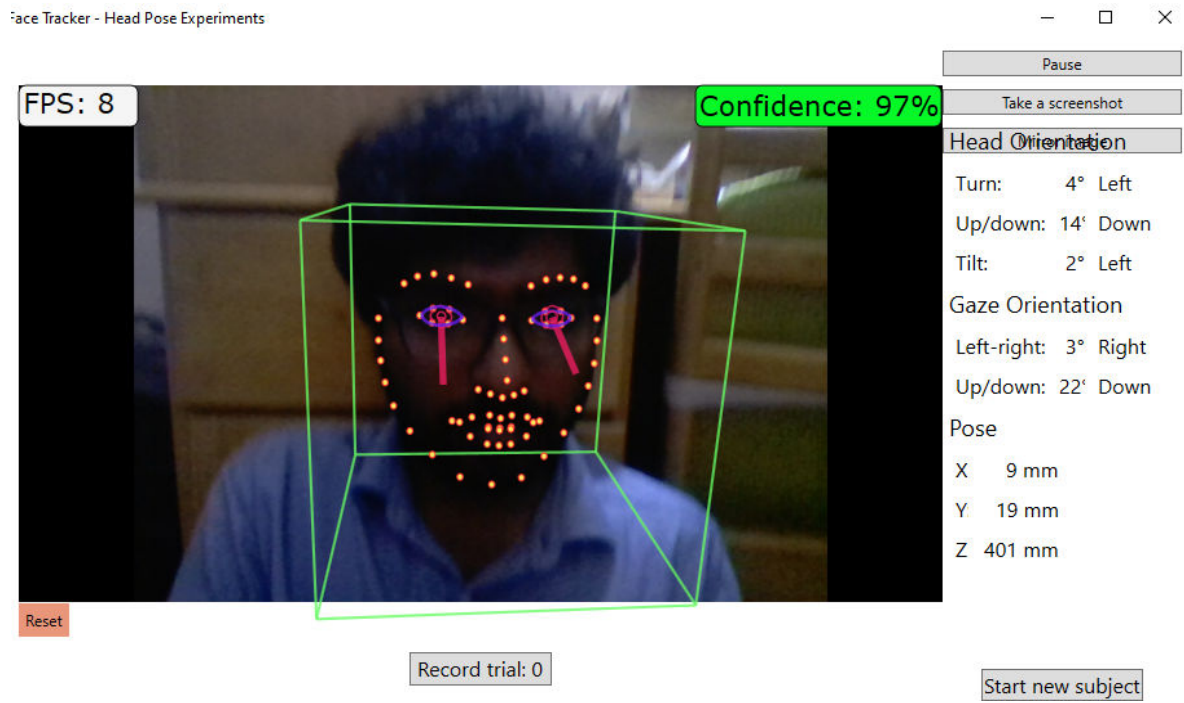

**Figure 3.** Screenshot from OpenFace Face Tracker

### 4.4. User Study

We compared all three eye gaze tracking algorithms through a user study involving both users with SSMI and their able-bodied counterpart. We wanted to use the eye tracker to operate a graphical user interface with limited number of screen elements, hence instead of traditional precision and accuracy measurement, we calculated the pointing and selection times for a set of fixed positions in screen.

**Participants:** We collected data from 12 participants – 6 were able bodied users (average age 28.4 yeas, 4 male, 2 female) and 6 users with SSMI (participants B, C, E, F, H, I).

**Material:** We used a Logitech C615 camera, with a resolution of 960 x 544. We used a 14.5" inch screen with a resolution of 1920x 1080. We measured the illumination on the user's face using a lux meter. For processing, we used an Intel NUC with i7 processor and 8 GB RAM.

**Design:** We created a user application in which we divided the screen into nine blocks and one of the blocks gets randomly highlighted with blue colour as shown in figure 4a. If the user clicks on the blue block, it turns green as shown in figure 4b and a different block was highlighted. If the user is unable to click on the highlighted block within 10 seconds, it turns randomly some other block to blue. Using this interface, we calculated the response time by measuring the time difference between appearance of a highlighted block and its selection. Able bodied users selected target using the left mouse button while users with SSMI selected target by dwelling on it for 500 msecs.

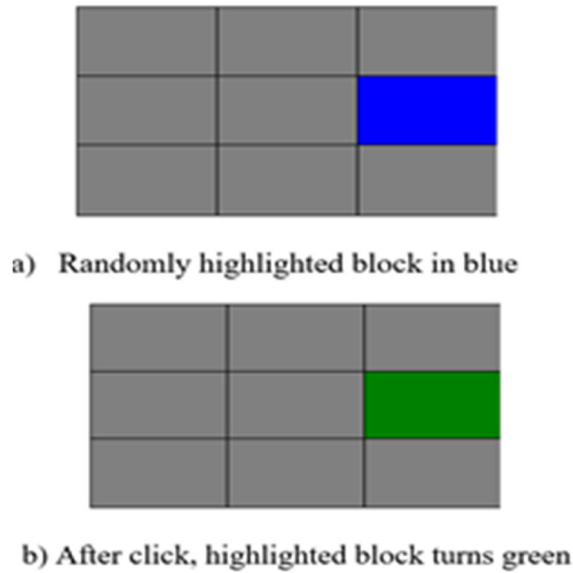

**Figure 4.** Pointing Task application

**Results:** We measured the time difference between onset of a target and its correct selection. On these pointing and selection times, we first undertook an unbalanced two-way ANOVA and found significant main effect of type of eye gaze tracker [$F(2,467)=98.76, p<0.01$] and interaction effect [$F(2,472)=12.87, p<0.01$]. Then we undertook two one-way ANOVA for both able bodied users and users with SSMI separately and found significant main effect of type of eye gaze tracker [$F(2,221)=57.51, p<0.01$ for ale-bodied users, $F(2,246)=48.15, p<0.01$ for users with SSMI]. The intelligent eye tracking system using OpenFace was significantly faster than the other two implementations (figure 5). Using the intelligent eye tracking system, able bodied users took 2.04 secs (stdev 1.66 secs) and users with SSMI took 3.28 secs (stdev 1.21 secs) on average to select target.

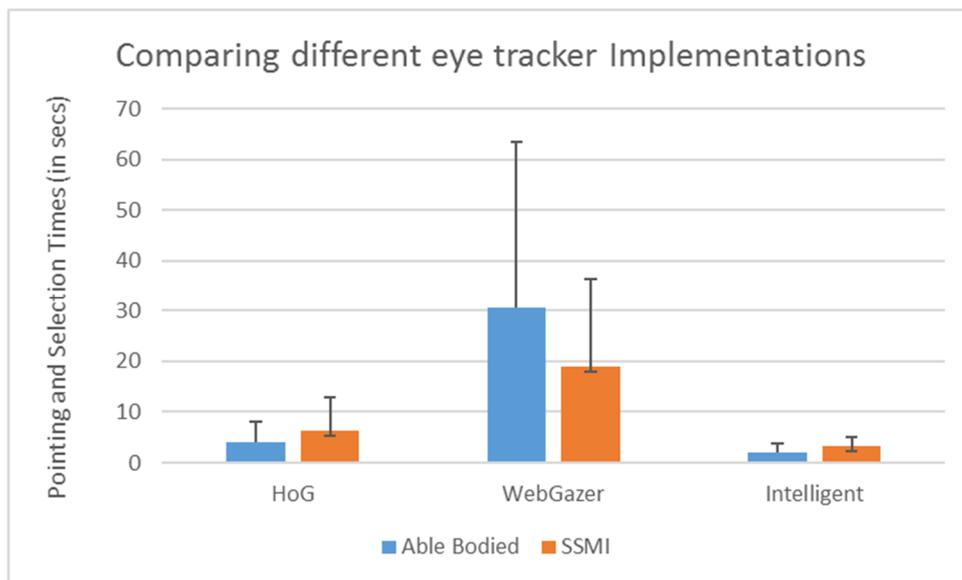

**Figure 5.** Comparing pointing and selection times among different eye gaze tracker implementations

**Discussion:** Our first approach for eye gaze estimation uses feature based approach. We used a method that extracts Histogram of Oriented Gradients (HoG) features combined with a linear SVM to detect eye landmarks. These landmarks were used to compute Eye Aspect Ratio (EAR) feature to estimate the gaze block on the screen. Even though HoG based landmark detection had been used earlier widely, we observed that it occasionally failed to detect landmarks of our users and affected the gaze estimation accuracy. This might be due to the fact that users with SSMI often have deformity in shape of head and their orientation of head is often different than their able-bodied counterpart. Variations in illumination and appearance of facial features like beard or spectacles could also affect tracking accuracy based on pre-selected facial features.

The second approach, Webgazer.js proposes to map the pixel data of eye images directly to gaze locations rather than to rely on handcrafted features from eye images. They used a 6 x 10 eye image patch for each eye and converted them to 120-dimensional feature vector. This vector is used as an input for a regression model to map to gaze points on screen. This approach also relies on multiple eye landmark detection algorithms which suffers similar limitations as HoG based algorithms. Further, this approach requires users to click at least 40-50 locations on screen for calibration purpose before it can make predictions which requires significant time.

OpenFace uses a state-of-the-art deep learning approach for landmark detection and gaze estimation. It uses Constrained Local Neural Field (CLNF) for eye landmark detection and tracking. Unlike HoG+Linear SVM approach, which was based on handcrafted features and trained on a relatively smaller dataset, OpenFace uses larger dataset and deep learning approach to learn the estimation of 3D gaze vector from the eye images. Even though OpenFace is not very accurate in predicting gaze points on screen, it does not suffer from illumination and appearance to detect eye landmarks. Further, OpenFace was implemented in C++ which makes real-time gaze estimation possible even on CPUs. In addition to these, the reported state-of-the-art cross-validation accuracy prompted us to test for a gaze block detection application. We used the OpenFace based gaze tracker for the user study reported in section 6.

## 5. Human Robot Interface

We developed a video see through eye gaze controlled human robot interface for users with severe speech and motor impairment to help in their rehabilitation process by undertaking different object manipulation tasks like painting and fabric printing. In particular, the fabric printing task involves a cooperation between an able-bodied person and a person with SSMI. The person with SSMI picks up a string about 6 cm in length and drop it on a powdered dye with his hand. He cannot pick it up himself from the dye. The able-bodied person picks up the colored string and rub it on a piece of paper. The whole process is repeated a few times with different colored powder dye. It may be noted the person with SSMI only chooses the color but cannot put the color at his or her chosen position on the piece of paper. We divided the task on two segments – a pick and drop task and a reachability task that can enable a person with SSMI to bring an object at any chosen position. In order to analyze the user acceptance of a relatively complex fabric printing task among users with SSMI, the pick and drop task and reachability study served as preliminary steps. Our proposed system allows a user to pick and drop an object by controlling a robotic arm based on his/ her eye gaze. The user can see the robotic arm and the object to be moved during interaction. In the following sections, we described the hardware and software setup, user interface and a set of kinematics equations to smoothly move the robotic arm in Cartesian space following users' eye gaze.

### 5.1. Proposed System

**Hardware Setup:** The proposed setup consists of a robotic arm, an eye tracker and a processing unit with a video see through display, which was implemented using a tablet computer. The robotic arm is made from a standard, off the shelf robotics kit. It consists of multiple linkages forming a simple chain, as shown in figure 6. It has 4 degrees of freedom, controlled by servo motors placed at the base, elbow and arm joints of the robot. Another servo motor is used to actuate the clamper mechanism at the front of the robotic arm. These servo motors are controlled by the ESP8266 based NodeMCU microcontroller. Information is sent to the microcontroller via serial communication through USB port of the tablet computer.

Initially, we have designed a task consisting of moving a set of objects from source locations to destination locations, as shown in figure 7. We used badminton shuttlecocks as our objects, due to the limited load capacity of the servo motors of the robotic arm. A subsequent reachability task involved moving a pen from a designated source to a randomly designated target point.

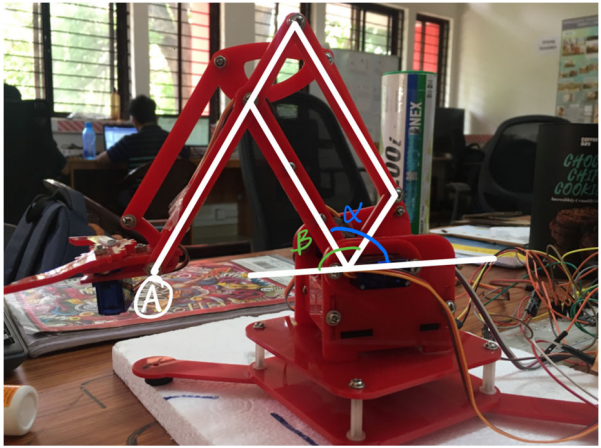 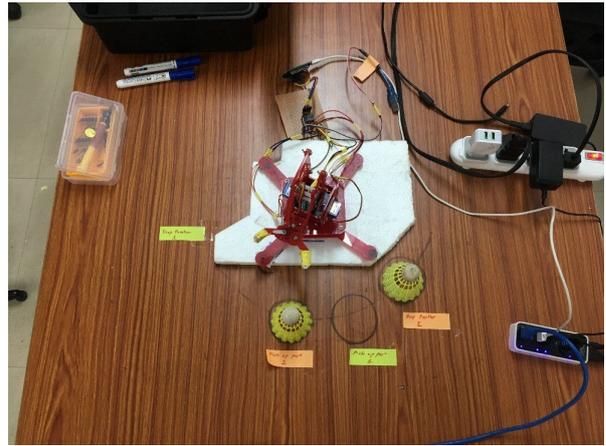

**Figure 6.** Linkages forming the robotic arm    **Figure 7.** Source and destination locations

**Software Setup:** The user interface for robotic control was transparent and overlaid on top of the live camera feed. It was setup as shown in figure 8, with the robotic arm placed behind the tablet computer. An eye tracker was placed at the bottom of the screen. The distance between the tablet and the robotic arm and camera viewing angle can be changed to accommodate both sources and destinations in the field of view of the camera. In our setup, the tablet was 1 foot away from the arm. We used low cost commercial off the shelf (COTS) screen mounted eye tracker and took the gaze coordinates from the tracker and feed them into our own algorithms, to smoothen and process gaze data. Our algorithm records gaze coordinates at 60 Hz and uses a median filter and Bezier curve to smoothly move a cursor based on eye gaze movement. During interaction, a red dot on the screen indicates the location of the user's eye at any given instant.

**UI for Pick and Drop Task:** The user interface for the pick and drop task only consists of transparent rectangles with green outline enclosing objects within the visual field of the video see through display. The transparent rectangles can be drawn at any position of screen and can be placed by detecting presence of specific objects using computer vision algorithms. The supplementary material shows that the robot can be trained to follow a blue cap within its working envelope. When a user dwells for a certain amount of time (500 milliseconds) in either of these regions, the object within the corresponding region is automatically picked up by the robotic arm, as shown in figure 8. After the object is picked up, the user was presented with another green rectangle, which indicates the position to drop the object (bottom right of figure 8a). On dwelling for 500 msecs in this region, the object is automatically dropped down as shown in figure 8b.

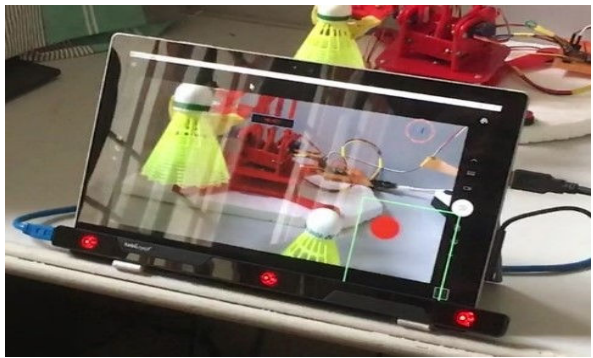 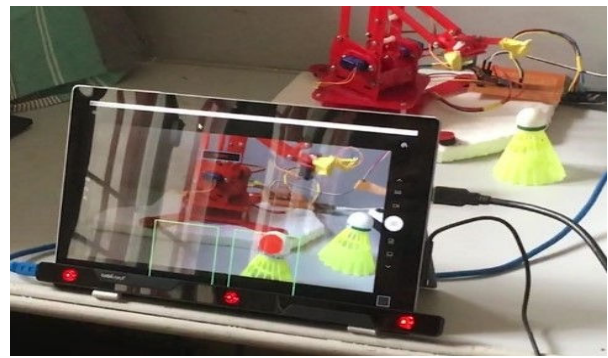

a. Robotic arm picking the object up    b. Robotic arm releasing the object

**Figure 8.** UI for pick and drop task

**UI for Reachability Task:** We designed another video see-through software with four-way and point to point control for moving the robotic arm at any random position within its field of reach. The UI consists of three screens, from the start screen, users can reach the point to point movement screen and bring the robotic arm at any point on the screen by clicking the mouse pointer on the screen. Users can also move the robotic arm in four directions by a fixed amplitude by pressing one of the four switches. The fixed amplitude of movement can be increased or decreased by pressing the buttons with plus and minus signs at the top right corner of the screen. It may be noted that the interface did not have any textual element and is accessible to different language speakers. Figure 9 shows a state chart diagram and figure 10 shows the UI elements of the system. The calibration procedure to map display coordinates to robotic coordinate system is explained in next section.

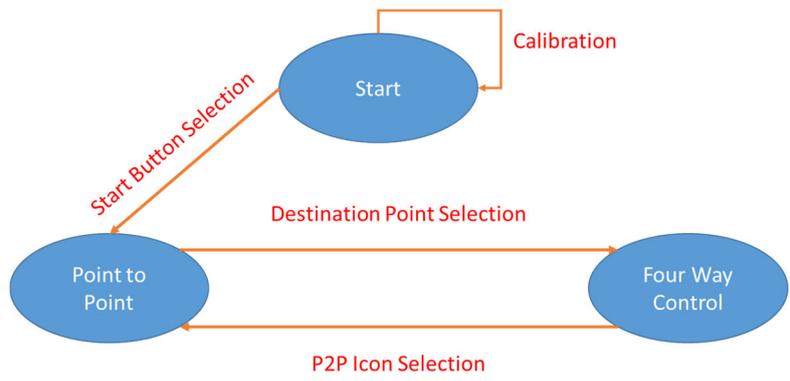

**Figure 9.** State chart diagram for the UI in reachability study

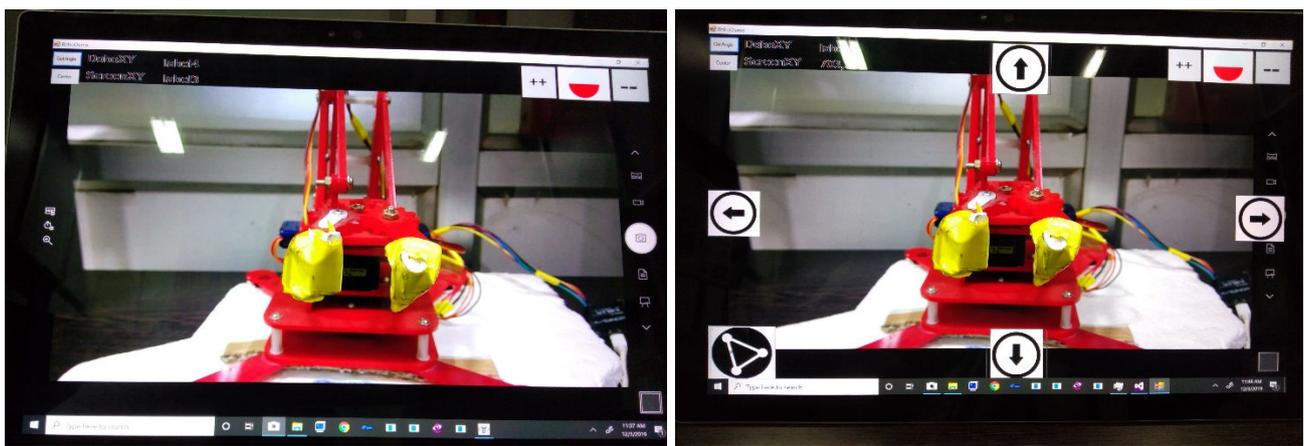

a. Point to point movement screen　　　　　　b. Four way control screen
**Figure 10.** Video see through display for four-way manipulation

**Kinematics Equation:** When users dwell at a particular position of the UI, the system had to send instructions to the robotic arm to reach that particular spot and then either grab or release an object. It required mapping display coordinates to a coordinates space of the robotic arm and mapping robotic joints into Cartesian space. The base joint of the robotic arm can make horizontal (left-right) movement but we developed following kinematics equations to control vertical motion (figure 11) of the robotic arm.

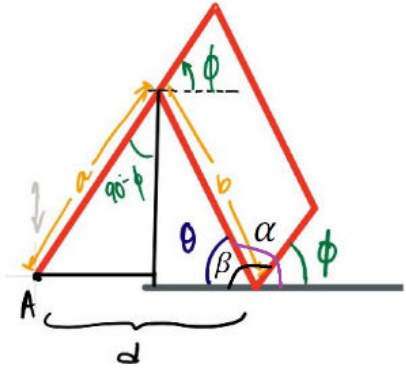 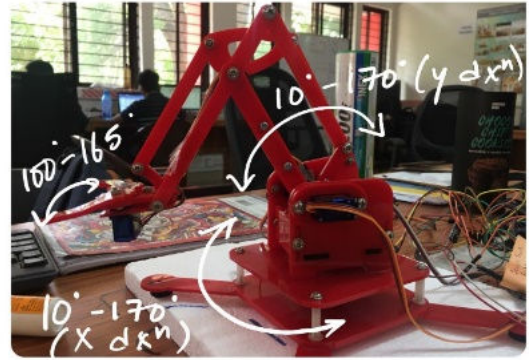

**Figure 11.** Different joints of the robotic arm

*Vertical Motion*

$\Phi = 180° - \beta$

$\theta = 180° - \alpha$

$d = a\sin(90° - \Phi) + b\cos(\theta)$
$\phantom{d} = a\cos(\Phi) + b\cos(\theta)$

*For constant $d$ (Vertical Motion of Point $A$)*

$\Delta d = 0$

$\Delta d = \Delta(a\cos(\Phi) + b\cos(\theta))$
$\phantom{\Delta d} = -a\sin(\Phi)\Delta\Phi - b\sin(\theta)\Delta\theta$
$\phantom{\Delta d} = 0$

$\Rightarrow \dfrac{\Delta\Phi}{\Delta\theta} = -\dfrac{b\sin(\theta)}{a\sin(\Phi)}$

Hence,

$\dfrac{\Delta\beta}{\Delta\alpha} = -\dfrac{b\sin(\alpha)}{a\sin(\beta)}$

To accommodate inaccuracies of the servo motors, we made one of the servos working till a certain angle after which the other one takes over. This improved the vertical range of the arm. In particular, when $\beta$ becomes 90 degrees, the motor controlling angle $\alpha$ kicks in and continues the motion from the following relation, the coefficients were calculated through empirical measurements.

$$\alpha = 184.275 - 0.438\beta$$

*Mapping display coordinates to the motion of the arm*

We used a calibration routine to map display coordinates to robot working space. The calibration routine involved manually dragging the robotic arm at a set of nine fixed positions within its working envelope. When the robotic arm reached a position, we recorded the coordinates of its gripper or far end of the arm in both display and robotic coordinate systems. A least square predictor based linear regression function was used to calculate coefficients of the mapping function. The $R^2$ value was 0.99 and the average RMS error was less than 1 cm for the experimental setup.

*Smoothing arm movements*

We smoothen the robotic arm movement as eye gaze movements occur at a higher frequency than supported by servo motors. This smoothing is not required for a high end robotic manipulator that has an upper limit in terms of acceleration and velocity profile. For the low cost arm, we converted a large movement into a bunch of smaller movements separated by a delay. The maximum angle which can be traversed before a delay was set at 30 degrees. The smoothing algorithm was as follows

- If difference in source and destination points is less than 30 degrees, no delay
- If difference is between 30 and 60 degrees, split into 2 motions and delay once
- If difference is more than 60 degrees, split into 3 motions and delay twice

We implemented both pick and drop and reachability tasks with a low cost robotic arm as well as a high end robotic manipulator, the 4-DoF Dobot Magician system from Shenzhen Technology Robotics. A demonstration of the final system can be viewed in the supplementary material.

### 5.2. User Studies

We undertook two user studies and a follow up study involving users with SSMI and their able-bodied counterpart to compare the performance of the system. We undertook two separate studies – in one we investigated picking up two objects and placing it at a fixed location. In the second study, users tried to move a robotic arm at a random position within its working environment. We described both studies in the next section following APA (American Psychological Association) format.

#### 5.2.1. Pick and Drop Task

**Participants:** We collected data from 9 able-bodied users and 9 users with motor impairment (participants A to I, except participants J and K). The able-bodied users were recruited from our university (7 males, 2 females, average age 23.5 years).

**Material:** We used a Tobii PCEyeMini tracker for eye tracking and a Windows surface pro tablet for rendering the user interface. We used a Kit4Curious robotic arm. It is a compact size, arduino compatible, 4 degree of freedom, smart robotic arm with 4 sg90 servo. It is a semi assembled robotic arm and can be re-assembled as per the design requirement. The arm is made by laser cutting smooth acrylic sheet. Outer dimensions are $22 \times 13.5 \times 13$ cm. The linkages are fixed using screws as pins so as to allow rotation about it. The load carrying capacity is limited to 10-20 grams due to low torque servo motors.

**Design:** Each user undertook two pick and drop tasks twice. We measured the time difference between initiation of the task and putting down the second object.

**Procedure:** We initially explained the aim of the trial to participants and to their teachers, parents and caretakers. After their consent, the eye tracker was calibrated for the participant. Then the experiment was explained to the participant and the participant was given a test run with the system. After this, the participant was asked to perform the task of picking and dropping both objects, while we recorded the time taken to complete the task.

**Results:** All 18 participants could undertake the trial on their first attempt and no participant reported any discomfort while using the system. Figure 12 shows the average and standard deviation of the duration for each trial and for each user group. As the data were not normally distributed, we undertook the Scheirer Ray Hare test as a non-parametric equivalent of two-way (Type of Participant × Session Number) ANOVA. We did not find any significant difference between participants $H(1,16)=2.94$, $p>0.05$.

Figure 13 plots the task completion time for each individual user separately. It may be noted that only 3 out of 18 occasions, the time duration of one pick and drop task exceeded 15 secs. It may be noted from figures 12 and 13 that, the task completion times did not change significantly between trial sessions and 5 out of 9 participants with SSMI took less time in second session while 4 took more time in second session.

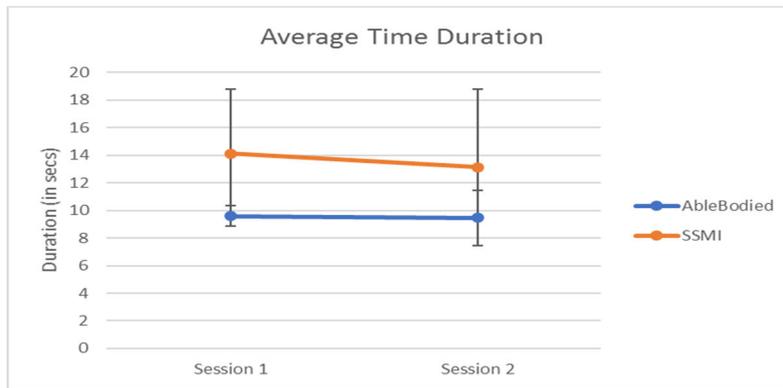

**Figure 12.** Total time duration of the task

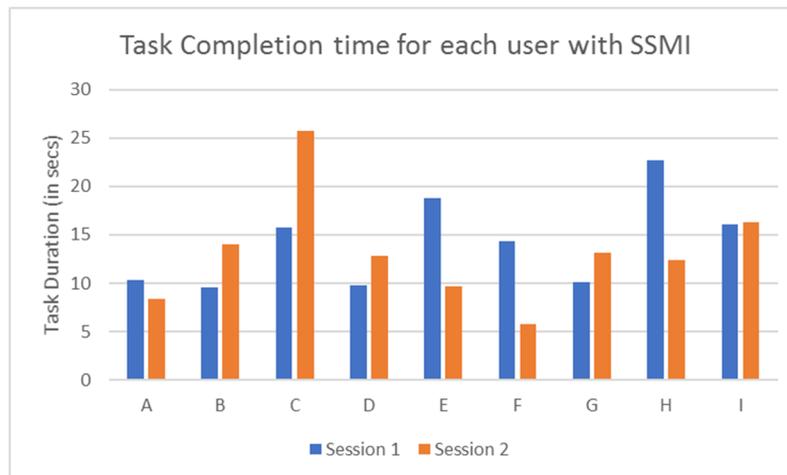

**Figure 13.** Task completion times for each user with SSMI

### 5.2.2. Reachability Study

In the previous study, we undertook pick and drop task from fixed locations. We undertook another study where users could bring the robotic arm at any random point within the field of reach of the robotic arm. However, in this study users only used the four-way control screen and did not use the point to point function as it was used in the previous study.

**Participants:** We collected data from 12 users- participants A, B, C, D, F, G took part in this study and we also collected data from 6 able-bodied users (4 male, 2 female, average age 28.2 years) at our university.

**Design:** For this study, we drew a target at a random positon in a sheet of paper and attached a pen with the robotic arm using a suitable gripper. The paper was half size of a A4 sheet (21 cm × 14.9 cm) and the target was drawn at a random position of the screen at a minimum distance of 5 cm away from the centre of the paper (figure 14). We instructed users to bring the pen at the designated target by using the four-way control (figure 10). The task was considered completed and stopped when the pen reached or crossed the innermost red circle. Like the previous study, users undertook pointing and selection using an eye gaze tracker. Selection was done by dwelling on target by 500 msecs. Each user undertook the trial twice. We recorded the path of the pen towards the target and the total task completion time. We also measured the number of times users changed direction of movement of the robotic arm. We followed similar procedure for the previous study.

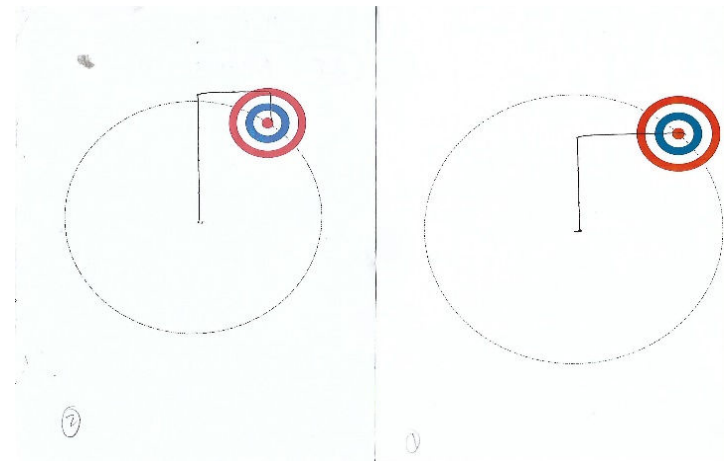

**Figure 14.** Task sheet for reachability study

**Results:** All 12 users could complete the task, which means they could reach the innermost red circle in both sessions. We undertook two-way ANOVA (*User × Session*) for the task completion time and number of times the arm direction was changed. We found

- Significant main effect of user for task completion time $F(1,5) = 14.36$, $p<0.05$, $\eta^2=0.70$
- Significant main effect of session for task completion time $F(1,5) = 9.44$, $p<0.05$, $\eta^2=0.68$
- Significant interaction effect of user and session for task completion time $F(1,5) = 6.51$, $p<0.05$, $\eta^2=0.68$
- Significant main effect of user for number of change in direction $F(1,5) = 49.7$, $p<0.05$, $\eta^2=0.91$
- Main effect of session and interaction effect of user and session was not found significant for number of direction changes

Figures 15 and 16 below shows task completion times and number of direction changes. It may be noted that able bodied users changed direction of movement more number of times than users with SSMI and similarly, at session 2, users with SSMI changed direction more number of times than in session 1. The task completion times were also less for able bodied users and at session 2 than session 1.

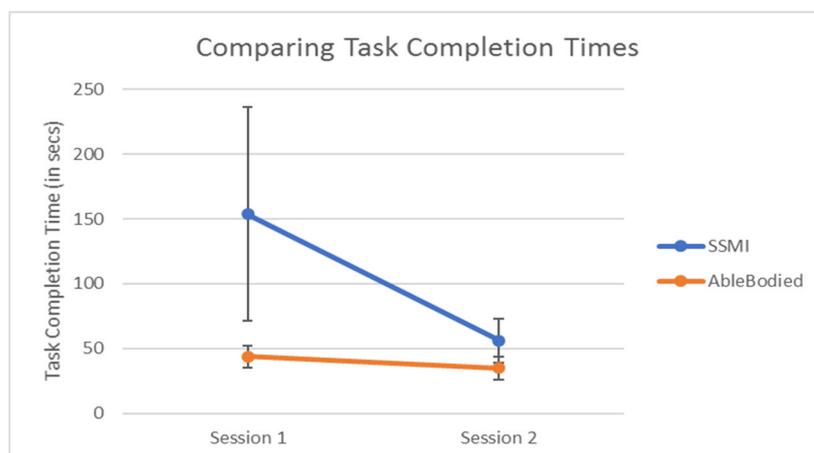

**Figure 15.** Comparing task completion times

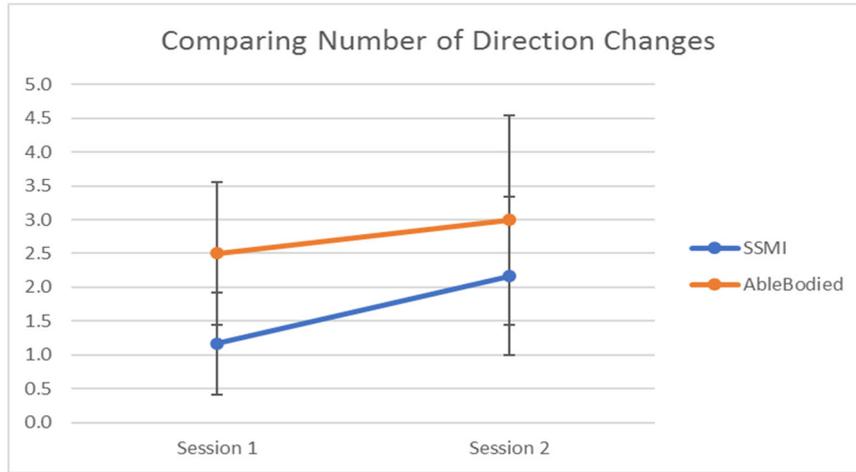

**Figure 16.** Comparing number of times of direction change

Figure 17 shows task completion times of each user with SSMI and it may be noted that 5 out 6 users improved interaction and reduced task completion time in the second session.

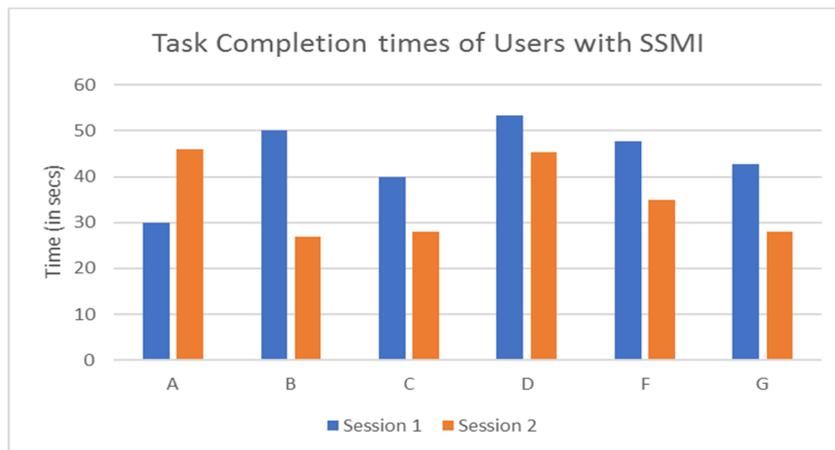

**Figure 17.** Task completion times for each user with SSMI

**Discussion:** Our work aims to help in the rehabilitation process of users with SSMI through use of cyber physical systems. We selected a particular task of fabric printing, which is currently undertaken by such users in a spastic society with the help of a caretaker. We developed and evaluated a proof-of-concept (PoC) eye gaze controlled robotic arm using low cost commercially off the shelf components. All nine users with motor impairment could use the system without assistance for two pick and drop tasks and the time duration, although higher than their able-bodied counterpart, but was not significantly different than able-bodied users.

In a second study, we investigated reachability of the robotic arm at a random position within the field of reach of the robot. Six users with SSMI could undertake the trial and reach the designated target within one minute in second session. We noted that users changed direction more number of times to reach target faster, which indicates selecting the same key multiple times by dwelling on the key was slower compared to selecting a different key. Future work will consider adapting the amplitude of movement based on users' previous interaction history, for example when a same key is chosen multiple times, we can progressively increase amplitude to reduce number of selections. Users with SSMI also required instructions in their native language and dialect while undertaking the task and that also contributed to slower interaction speed. Following, Zaira's [30] work, we are also investigating different velocity and acceleration profiles in the kinematics equations. Finally, it may be noted that our users used this system for first time and we can add more screen elements for point to point movement, for example, we can use eight-way control instead of four-way control to increase interaction speed further.

## 6. Webcam based gaze controlled AR HRI

The previous two sections described development and evaluation of webcam based eye gaze tracker and video see through human robot interface. As both activities went on in parallel, we used commercial eye gaze trackers for initial evaluation of the human robot interface. In the following section we have described a study involving users with SSMI that controls the robotic arm using the webcam based gaze tracker and video see through display.

### 6.1. User Study

This study was designed similar to the reachability study described in the previous section, but users were allowed to make both point-to-point and jog movement using the four-way control. We used similar set of material used in the previous studies except the COTS eye tracker.

We undertook the trial with 6 users with SSMI (participant codes A, B, D, F, H, I). We ensured all these participants took part in at least one of the HRI studies described in previous sections. All 6 participants were able to successfully complete the task by reaching the target. Out of these 6 people, two users undertook the trial twice and one user did it three times. The average time of completion of task for the participants was 2.45 mins with minimum and maximum time being 1.7 and 3.44 mins respectively. It may be noted from trendlines of figure 18 that for all three users who undertook the task more than once, the completion time is reduced with practice.

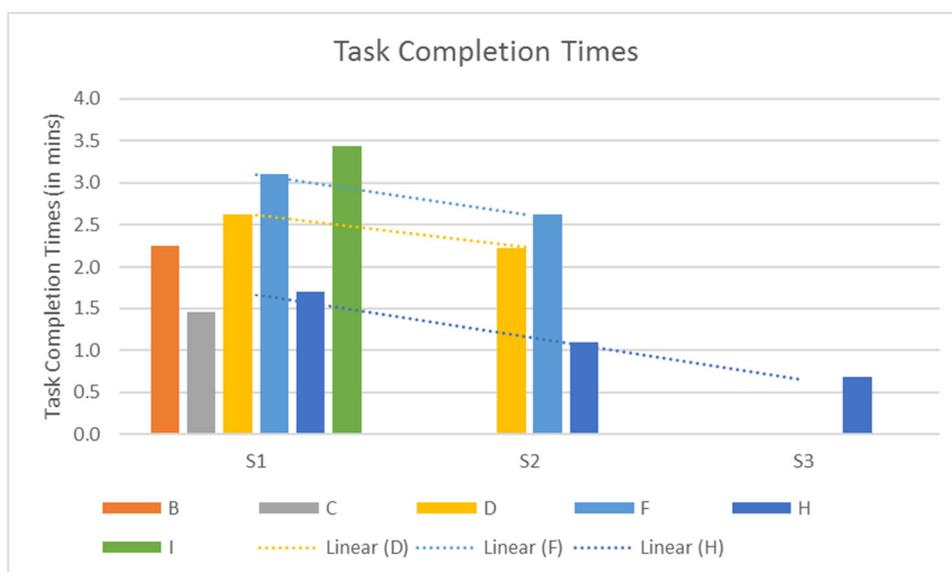

**Figure 18.** Comparing tasks completion times of the HRI task with webcam based eye gaze tracker

A video demonstration of the system with narration can be found at https://cambum.net/CEFC/ARGazeRobot.mp4

### 6.2. Discussion

Many earlier research explored controlling robotic manipulator or wheelchair through eye gaze or head movement although many of them were not evaluated with more than 5 users with severe motor impairment. We worked with a set of users who are severely motor impaired and the level of disability is more severe than participants recruited for earlier research works. We undertook two parallel activities on developing a webcam based eye gaze tracker that does not require procuring any extra hardware with a standard laptop and developing a video see through human robot interface that does not require users to wear any eye tracking glass. The video see through interface only used icons and thus could be used by different language speakers. Our user studies demonstrated that users with SSMI can undertook representative pick and drop tasks using our system. We noted that the average task completion times were higher for webcam based gaze tracker than COTS eye gaze tracker. A COTS eye tracker uses dedicated ASIC chip for video processing while a webcam based gaze tracker uses the computer processor, which is shared with other programs. It may also be noted that we did not use any high end GPU (graphical processing unit) for the reported trials. An earlier study [Shree 2019] undertook detailed analysis on visual search and fixation patterns on similar set of users and found that they also suffer from Nystagmus and cannot fixate attention in all regions of screen with equal ease as their able bodied counterpart. However, even considering these difficulties it may be noted that with practice, the task completion times can be reduced further and will be nearly equal to COTS eye gaze tracker for a user interface with limited screen elements. In the following sections, we added further justification on design of user studies and value addition.

**Use of COTS eye gaze tracker:** Development of the video see through display based robotic manipulator and webcam based eye gaze tracker went in parallel. Even the latest version of webcam based gaze tracker is not as accurate as the COTS Tobii tracker. Initially, we used the Tobii tracker as the webcam based gaze tracker was not ready and also to train participants with an accurate tracker for HRI tasks. After trained with the video see through interface with the COTS eye gaze tracker, users found it easy to use the webcam based less accurate tracker for the same task.

**Choice of robotic manipulator:** We initially used a low-cost robotic arm for the study. The advantages of the low cost system were our end users could buy it for personal use and it was easily scalable as the components can be 3D printed in large volume. However, the system was not robust, the kinematics equation required adjustment through empirical measurements due to inaccuracies in servo motors and it has less than 50 gm payload. Our end users never used any robotic manipulator before and so we started the research with a low cost system. After getting promising results, we integrated our software with a 4-DoF robotic system (Dobot Magician) from Shenzhen Technology Robotics. The advantages of the Dobot system were it was robust, the software development kit already implemented kinematics equations and we could directly control it in Cartesian space. However, it was costlier than the previous system and our end user could not afford it personally. We implemented both pick and drop and reachability tasks with both low cost robotic arm and Dobot systems. From interfacing point of view, the low cost robotic arm required direct electronic interfacing and we used NodeMCU wireless units to send commands to servo motors. For the Shenzhen Dobot, we can use the SDK and send software commands to the robot. Interfacing at the software level is more reliable and easier to maintain than sending commands electronically to individual servo motors. In both cases, the software user interface remains same and abstracts the underlying electronic connection and hardware manipulations from users. The same interface can also be extended to UGV or UAV through a multimodal joystick controller as described in the next sub-section.

**Multimodal Joystick Controller:** We have invented a generic joystick controller mechanism (Indian Patent Application no. 201941044740) that can be operated by multiple modalities like speech, gesture, eye gaze and so on. It is a scalable approach as it does not require access to the hardware or software of the controlled device like drone, AGV or robot. Using this device any commercial-off-the-shelf (COTS) cyber physical system can be operated through multiple modalities without interfering into its hardware or software, even if it does not have a software development kit (SDK).

The hardware part of the mechanism consists of an assembly to move the analog sticks in desired direction and press buttons. The stick has two U-shaped components – one component is attached to a joystick and mounted to the other U-shaped component (figure 19). A servo motor is attached to the system to move the first component (figure 20). The sticks on the console are controlled by a mechanical linkage run by suitable servo motor. This mechanical linkage can also be modified to fit all the shape and sizes of Joystick controllers available in the market. It has curved end to fit on the servo motor shaft. The rectangular cavity slides in the stem of the joystick. The whole mechanism can by mounted on the console itself of can be independently placed in the close vicinity.

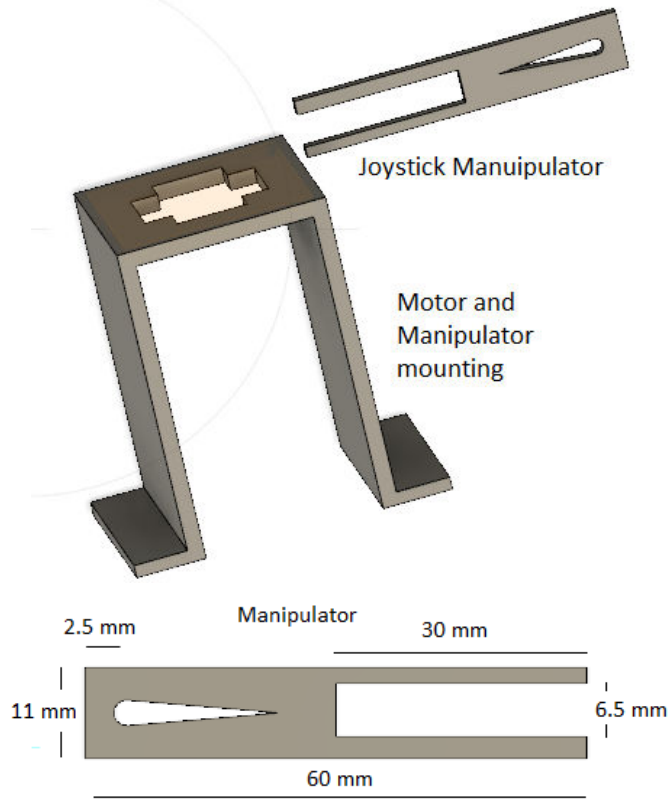

**Figure 19.** Picture of the components of the Joystick Controller

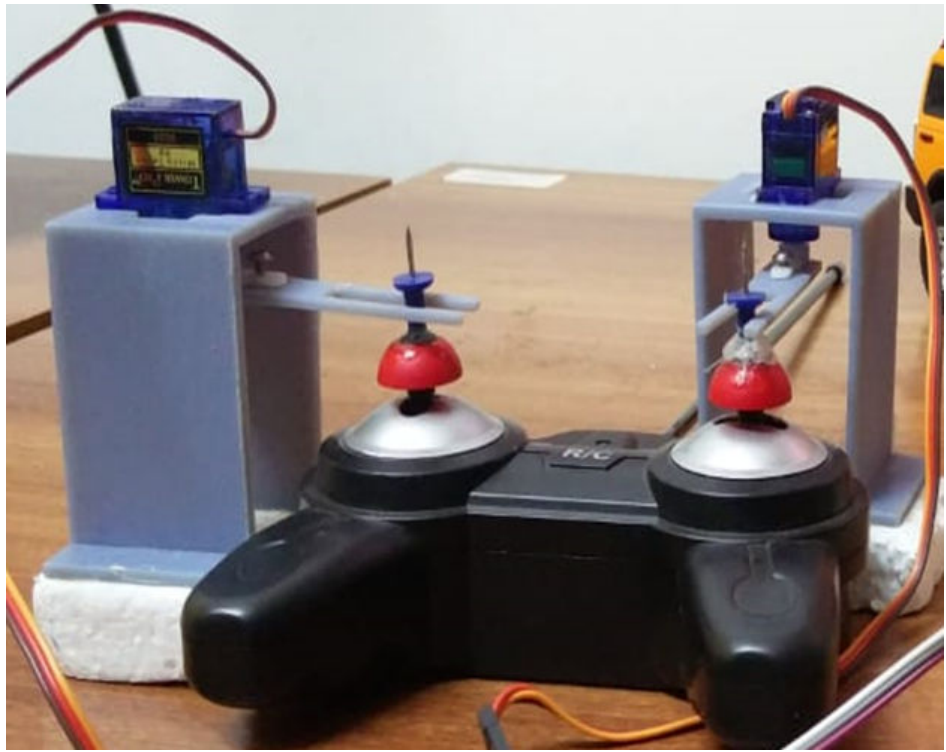

**Figure 20.** Mounting of Servo Motor

The electronic part of the system was developed using a Node MCU WiFi module that communicates between a laptop or PC and the joystick controller. The circuit diagram is furnished below in figure 3.

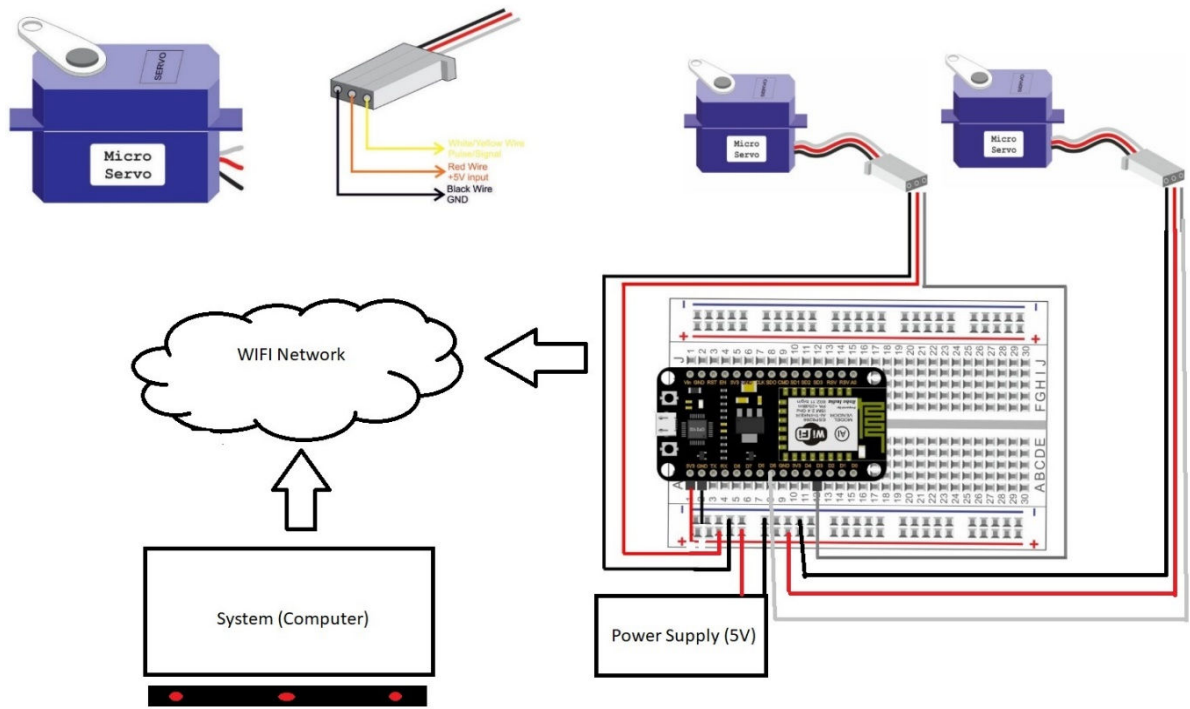

**Figure 21.** Electronic Circuit Diagram

The software part of the invention consists of a multimodal module consisting of eye gaze tracking, gesture, speech recognition systems and live video feed from a camera. A software module sets up a precedence among different modalities and send signal to the electronic part to control the joystick controller.

Once we have the mechanism in place, we need not move the Joystick manually. We can use different modalities to operate the joystick like Keyboard, Eye gaze, Gesture, Speech etc. These modalities take leverage of the mechanical mechanism and allows the users to interact with car, drone and gaming devices in a more involving and responsive manner. It is like a mechanical hand operating the Joystick controller for us, while we instruct the mechanism through keyboard, gesture, eye-gaze, speech and so on. The user gets a nice feedback and the mechanism is very responsive. The mechanism also can be used by persons with disability, motor impairment and situationally impaired as it does not require them to move hands or press buttons. So this mechanism tried to bridge gap between those who wish to move around and do things but cannot because of physical constraints. A video demonstration can be found in the supplementary video from 41 secs onwards and by learning only one user interface, users will be able to control robotic manipulator, UGV or UAV.

**Why two studies for HRI:** In human computer interaction, a plethora of research has been conducted on rapid aiming movement citing pioneering studies by Woodworth [1899] and Fitts[1954], which established existence of two phases of movement for any reaching or fetching task– a ballistic phase and a homing phase. Our studies roughly correspond to separately investigating the homing and ballistic phases of movement. The first study did not require manipulating movement of the robotic arm, rather focusing attention on a fixed source and destination for pick and drop. The second study only involved manipulating the robotic arm with a four-way control without any task on a separate object manipulation. We analysed the pattern in manipulation using four-way control and pointed out modifications to improve interaction speed.

**Value addition:** Alsharif [2016] reported lack in research for eye gaze controlled HRI while Palinko [2016] reported higher efficiency in eye gaze controlled system compared to head controlled HRI. However, we did not find many work on HRI for people with severe motor impairment. Alsharif's [2016] study involved only one participant with motor impairment while Zaira [2014] reported similar four-way control for robotic prosthesis manipulation through eye gaze but did not report range of abilities of participants. Our study compared representative object manipulation tasks between users with SSMI with their able bodied counterpart. It may be noted that our end users could not make hand gesture or synchronously press a switch to select like Kim's [2001] study. As part of the work we proposed a gaze controlled video see through display, which is less invasive than an eye tracking glass or head mounted system [Kim 2001, Kohlbecher 2012]. The user interface did not use any textual label and can be usable by different language speakers. Our present users except participant H never manipulated any physical object themselves and through our system they could manipulate an object first time in their life. Although our present system reported slower interaction speed for users with SSMI than their able bodied counterpart but future work is investigating to reduce this digital divide further not only for robotic arm but also for other cyber physical systems.

**Future work:** Our future work will design a gripper for the robotic arm for undertaking the fabric painting task and deploy the system with the same user interface. We have integrated the gaze controlled video see through display with a robotic arm with higher payload and a short clip can be found at the supplementary material. It may also be noted that the gaze controlled see through interface can be used for other cyber physical systems like robotic wheelchair, UAV or UGV besides the robotic arm. In parallel we are working on increasing the accuracy of the webcam based gaze tracker and creating a dataset involving users with SSMI.

## 7. Conclusion

This paper presents an eye gaze controlled robotic arm to help in the rehabilitation process of users with severe speech and motor impairment. Initially, we compared different algorithms to estimate eye gaze from webcam and compared their performance through user studies. Using the OpenFace based intelligent eye gaze tracking system, users with SSMI can select one of nine regions of screen in 3 secs on average. We developed new algorithm to control a video see through graphical user interface (GUI) with eye gaze, use the GUI to control a robotic arm and kinematics equations to smoothen the arm movement for pick and drop task. The system was evaluated with 18 users (9 able-bodied, 9 users with SSMI). All users could undertake the designated task twice in first attempt in an average time less than 15 secs. A second study undertook a reachability task and 12 users (6 able-bodied, 6 users with SSMI) could bring the robotic arm at any random position within its field of reach in less than 60 secs. Finally, we evaluated the whole system involving the webcam based eye gaze tracker and video see through human robot interface through a user study involving people with SSMI and they could bring the robotic arm at a pre-designated point within its working envelop in 2 mins on average. Their performance improved while they under took the task second or third time than the first time  Future work is improving the robotic control algorithm to further reduce task completion times for representative object manipulation tasks.